\documentclass[twocolumn,prl,aps]{revtex4}

\usepackage{epsfig}
\usepackage{graphicx}
\newcommand{\ket}[1]{| #1 \rangle}
\newcommand{\bra}[1]{\langle #1 |}

\begin{document}

\author{R. H. Koch, J. R. Rozen, G. A. Keefe, F. M. Milliken, C. C. Tsuei, J. R.
Kirtley, and D. P. DiVincenzo}

\title{Low-bandwidth control scheme for an oscillator stabilized Josephson qubit}

\affiliation{{IBM Watson Research Ctr., Yorktown Heights, NY 10598
USA}}

\date{\today}

\begin{abstract}
We introduce a new Josephson junction circuit for which quantum
operations are realized by low-bandwidth, nearly adiabatic
magnetic-flux pulses. Coupling to the fundamental mode of a
superconducting transmission line permits a stabilization of the
rotation angle of the quantum operation against flux noise.  A
complete scheme for one-qubit rotations, and high-visibility
Ramsey-fringe oscillations, is given.  We show that high
visibility depends on passing through a portal in the space of
applied fluxes, where the width of the portal is proportional to
the ramp-up rate of the flux pulse.
\end{abstract}

\maketitle

Among the many candidates for the physical implementation of a
quantum computer, Josephson junction circuits have always been
among the most promising.  The quantum behavior of these circuits
is readily tailorable by the choice of electrical topology and
circuit parameters: there are various regions of this parameter
space in which a coherent, controllable two-level quantum system,
suitable for the realization of a qubit, is possible.  This same
tailorability must also be exploited to avoid strong coupling to
the environment and other decohering effects. The complexity of
this optimization\cite{Devrev} is such that many distinct
Josephson circuit qubits are under active study, with successful
single qubit control and two-qubit coupling achieved in a number
of cases\cite{lots,vion}.  But further improvements in qubit
performance are unquestionably needed and continue to be sought.
Recently, another important degree of freedom has been added to
this search: strong, coherent coupling between a Josephson
junction qubit and a quantized harmonic oscillator, realized by a
superconducting transmission line, has been
achieved\cite{Yale,Delft}. This exciting discovery raises the
question of how best such a coupled system is to be exploited for
quantum information processing.

In this Letter we report a new class of Josephson flux qubits that
enter novel regions of the design space to achieve superior qubit
performance.  Here are its features:  First, our qubit can be
placed in a ``frozen" state in which the barrier is very high,
which makes resetting and measuring the qubit very reliable.
Second, and more important, all qubit operations are realized by
low-bandwidth, nearly adiabatic operations.  Since the amount of
environmental noise seen by the qubit is proportional to this
bandwidth, we gain significantly by requiring only an
approximately 1 GHz control bandwidth rather than the many GHz
that are necessary in other control schemes, which require the
transmission of microwave radiation to the qubit. Finally, within
our adiabatic operation scheme, the presence of a coupled quantum
harmonic oscillator plays a specific role in the qubit operation:
the qubit rotation angle, which is equal to the dynamical phase
difference accumulated between ground and excited state energy
surfaces, is stabilized by the adiabatic conversion of the qubit
states to the ground and excited states of the quantum oscillator.

The resulting qubit with its associated control transmission lines
is complex, but the advantages gained by our operation strategy
leads to a qubit for which a scale up to larger systems will
eventually be more feasible.  Our experimental realization of this
qubit system is shown in Fig. 1.  The device consists of three
Al/Al$_2$O$_3$/Al Josephson junctions grown using an in-house
shadow mask process, arranged in a gradiometer pattern. This
design is a modification of a qubit previously reported\cite{BKD}.
The body of the gradiometer is aluminum and consists of three
loops. The lower loop is threaded by external flux $\Phi$, the
small left loop is threaded by the flux $\Phi_C$; the upper loop,
the ``pickup" loop, is inductively coupled to a high-Q niobium
superconducting microstrip ``pickup" transmission line. There are
no wires attached to the qubit.  Readout is done using a dc SQUID
inductively coupled to the pickup transmission line. The qubit is
operated at 30 mK; however the effective electrical temperature
from the circuits that drive the fluxes $\Phi$ and $\Phi_C$ is
about 1.3 K.  These circuits are not shown; however, these are
coupled to the qubit with two additional superconducting
microstrip transmission lines.  The lines are shorted at the end
that is not connected to the room temperature electronics. The
length of the lines between the qubit and the shorted ends are
such that at the desired frequency of operation of the qubit, 1.54
GHz, there will be a current node at the location of the qubit.
Hence, when we operate the qubit at the degenerate point, which we
can tune to have an operating frequency of 1.54 GHz, the
contributions from both the low-frequency noise and the noise at
the operating frequency are minimized.

\begin{figure}[htb]
\includegraphics[width=3.25in]{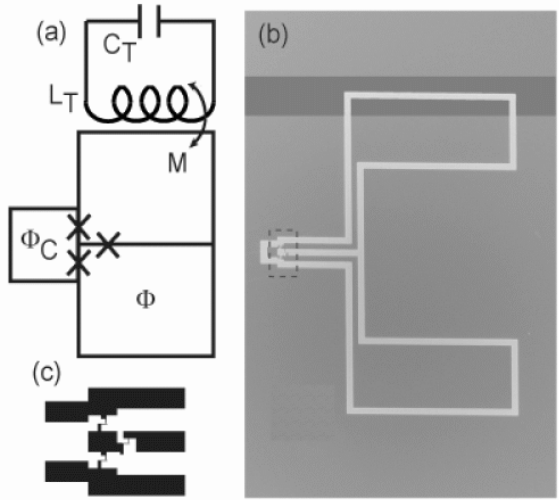}
\caption{(a): Schematic layout of the qubit and the harmonic
oscillator equivalent of the transmission line.  (b): Picture of
the qubit.  The qubit is 500 $\mu$m long and each junction is 250
by 250 nm$^2$ in size. On the upper loop the pickup transmission
line is shown. (c): Mask art of the three shadow type Josephson
junctions (blow up of dashed box in (b)).}
\label{fig1}
\end{figure}

Using our network graph theory\cite{BKD}, we have obtained a
comprehensive quantum mechanical model for this qubit.  The system
Hamiltonian is equivalent to that of a massive particle in a
four-dimensional potential.  The particle mass is set by the
system capacitances; the four degrees of freedom are the three
Josephson junction phases plus the phase (the time integral of a
voltage) across the capacitance of the transmission line LC
resonance.  Other environmental degrees of freedom, e.g., those
associated with the control transmission lines, are modelled as in
\cite{BKD} using an oscillator bath.  We calculate that the
intrinsic quantum lifetimes associated with this bath, for the
choices of control parameters described below, and for $T=100$ mK,
are $T_2=600$ nsec and $T_1=300$ nsec, which are more than long
enough for our adiabatic control scheme to successfully execute
quantum operations. In practice the coherence of our qubit is
further decreased by extrinsic noise in the control electronics,
an example of which will be analyzed below.

While the resulting potential function is complex, there is a
region of two-dimensional parameter space of applied fluxes $\Phi$
and $\Phi_C$ (see Fig. 1) in which the potential has a simple,
double-well structure. The lowest energy states of this double
well form the qubit; the qubit is also linearly coupled to the
transmission line harmonic oscillator, as we will describe
shortly.

In the vicinity of these double-well minima, the three degrees of
freedom of the qubit can be linearly transformed into two ``fast"
degrees of freedom transverse to the double-well axis and one slow
longitudinal phase which we will call $\delta$. Using a
Born-Oppenheimer approximation\cite{Bruder} for the fast degrees
of freedom, we have numerically obtained an effective
one-dimensional double-well potential $V_Q(\delta)$ for the qubit.
While we use the full numerical form in our calculations, it is
useful for describing features of this potential to represent it
using the following simple analytic anharmonic form\cite{foot1}:
\begin{equation}
V_Q(\delta,t)=-h_2(t)\delta^2+h_4\delta^4+a(t)\delta.
\label{smodel}
\end{equation}
We indicate here an explicit dependence on time $t$, as our qubit
control technique involves the pulsing of $\Phi$ and $\Phi_C$ in
time.  In our devices, to good approximation, $\Phi$ controls the
double-well asymmetry coefficient $a$ and $\Phi_C$ varies the
double-well barrier height $h_2$. We find that $h_2$ can be either
positive or negative; that is, the double-well structure can be
transformed into a single well.

Eq. (\ref{smodel}) is only one part of the description of the
system; coupling to the harmonic-oscillator phase $\varphi$ must
also be included.  To good approximation, the full Hamiltonian of
the system can be written
\begin{eqnarray}
H&=&{e^2Q^2\over 2C_Q}+V_Q(\delta,t)+\label{firstjc}\\&
&{e^2q^2\over 2C_T}+ \left ({\Phi_0\over 2\pi}\right
)^2{\varphi^2\over L_T}+ \left ({\Phi_0\over 2\pi}\right
)^2{M\,\delta\cdot\varphi\over L_QL_T}.\nonumber
\end{eqnarray}
Here the first two terms are the qubit Hamiltonian, the next two
terms are the harmonic oscillator Hamiltonian, and the last term
is the linear coupling between the two.  $Q$ and $q$, the qubit
and oscillator charge operators, act as momentum operators in the
Schrodinger equation: $[\varphi,q]=[\delta,Q]=i$.  $C_Q$ is the
capacitance associated with the Josephson junctions; in our
circuit all three have $C_Q\approx 50fF$ (there is substantial
excess capacitance in parallel with the intrinsic capacitance of
the oxide junctions).  $C_T\approx 1fF$ and $L_T\approx 10nH$ are
the effective transmission line capacitance and inductance,
$L_Q\approx 640pH$ is an effective qubit inductance (actually a
complicated function of the full inductance matrix of the qubit
circuit\cite{BKD}), $M\approx 50pH$ is the mutual inductance
between the qubit and the transmission line, and $\Phi_0=h/2e$ is
the superconducting flux quantum.  In other studies\cite{Yale} the
physics of this situation has been described using a
Jaynes-Cummings Hamiltonian; we find that retaining the
first-quantized form of Eq. (\ref{firstjc}) has proved helpful in
analyzing the details of our control scheme.

In our device, the barrier height $h_2$ in the model potential Eq.
(\ref{smodel}), at low values of the control flux $\Phi_C$, can
easily be made so high that the quantum tunnelling rate $\Delta$
can be made very small ($\hbar/\Delta$ being many minutes at
least). In this regime the qubit state is essentially frozen, with
two degenerate eigenstates $\ket{L}$ and $\ket{R}$ for ``left" and
``right" well, see Fig. 3, point A.  The device is hysteretic in
this regime, so it can be stably set in either $\ket{L}$ or
$\ket{R}$ by a suitable quasistatic sweeping of flux $\Phi$. In
addition, the state is easily read out by a sensing SQUID
inductively coupled to the transmission line of Fig. 1.

So, for $\Phi_C=\Phi=0$, corresponding to $a,\Delta\approx 0$, the
qubit is in a ``resting condition" in which its dynamics are
frozen.  Starting from this resting condition, conceptually there
is a simple prescription for adiabatically applying single-qubit
quantum gates to the qubit.  A ``$Z$ rotation" (of the Bloch
sphere defined by the $\ket{L,R}$ basis) is simply obtained by
adiabatically varying $a$ in Eq. (\ref{smodel}) (by pulsing the
main-loop flux $\Phi$) with $\Delta$ held equal to zero. In the
adiabatic limit, the $Z$ rotation angle is given by $\phi=2\int dt
\epsilon(t)/\hbar$, where
$2\epsilon\equiv(\bra{L}V_Q\ket{L}-\bra{R}V_Q\ket{R})$ is the
difference in depths of the two wells of $V_Q$.

The ``$X$ rotation" scheme will require more discussion, as it
involves a stabilizing effect due to the harmonic oscillator
degree of freedom in Eq. (\ref{firstjc}).  If, starting from the
resting state, the control flux $\Phi_C$ is pulsed, then the
$\epsilon$ defined in the last paragraph will remain zero and
$h_2$ will be varied. Since ideally the double-well potential Eq.
(\ref{smodel}) is symmetric throughout this pulse, the
lowest-lying eigenstates of this potential will always be
symmetric ($\ket{S}$) and antisymmetric ($\ket{A}$) with respect
to the qubit coordinate $\delta$.  In the resting state
$\ket{S,A}=(\ket{L}\pm\ket{R})/\sqrt{2}$; these basis states lie
at right angles to the $\ket{L,R}$ states on the Bloch sphere, so
that adiabatic evolution along this coordinate and back results in
an $X$ rotation by angle $\theta=\int dt \Delta(t)/\hbar$, where
$\Delta(t)=E_1(\Phi_C(t))-E_0(\Phi_C(t))$, and $E_{0,1}$ are the
first two instantaneous eigenenergies (see Fig. 2) of the qubit
Hamiltonian Eq. (\ref{firstjc}).

\begin{figure}[htb]
\includegraphics[clip,width=8.5cm,trim=85 180 100 200]{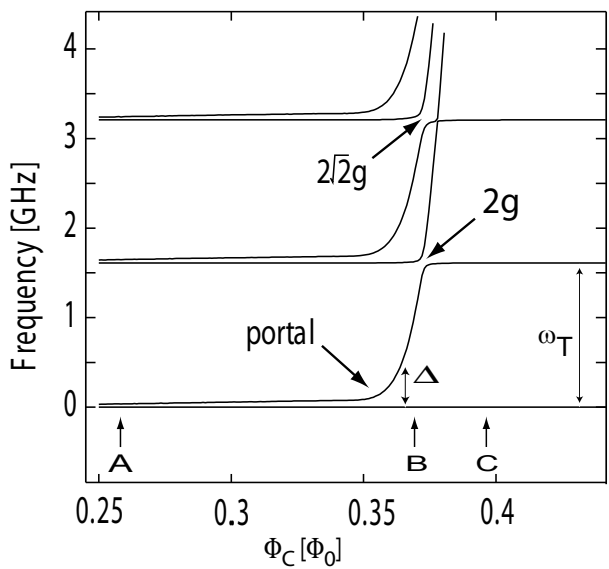}
\caption{Instantaneous eigenvalues of $H$, Eq.
(\protect\ref{firstjc}), as a function of control flux $\Phi_C$,
for main flux $\Phi\approx 0$.  The exponentially rising tunnel
splitting $\Delta$, and the $\Phi_C$-independent oscillator
frequency $\omega_T$, are indicated.  At $A$, the ``resting
state", the double-well barrier is very high and the two lowest
eigenstates are nearly degenerate; here the resting state is not
quite symmetric, with $\epsilon\approx 15$MHz.  For this nonideal
case successful quantum operation requires that the system pass
through the "portal" as described in the text.  At point B, beyond
the portal, quantum tunneling dominates the dynamics and the
system is effectively symmetric.  Beyond B the qubit comes into
resonance with the transmission line oscillator, with a splitting
of $2g\approx 220$MHz. At point $C$, the lowest lying states have
purely harmonic oscillator character. } \label{fig2}
\end{figure}

\begin{figure}[htb]
\includegraphics[clip,width=8.5cm,trim=50 150 50 150]{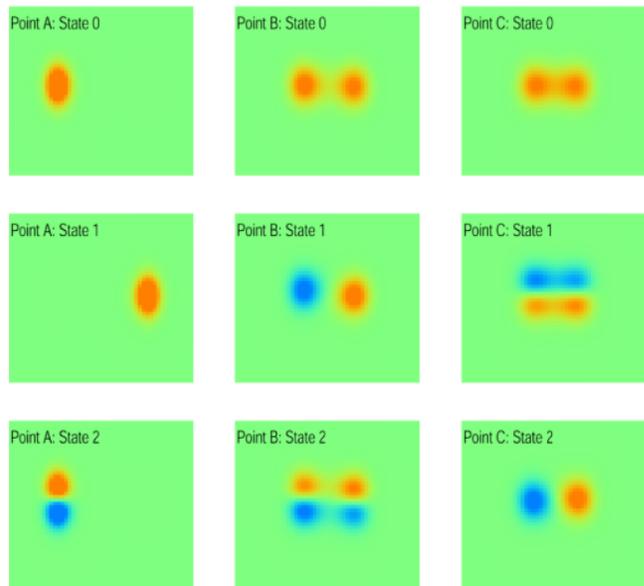}
\caption{Color density plot of first three first energy
eigenstates of $H$ at points $A$, $B$, and $C$ of Fig. 2. $\delta$
is the horizontal axis, and $\varphi$ the vertical axis (see Eq.
(\protect\ref{firstjc})). For $A$ the first two eigenstates,
localized in the left and right wells, are not quite degenerate
because $\epsilon\neq 0$ (see Fig. 2).  In going from $A$ to $B$,
on account of strong quantum tunneling, the first two eigenstates
change rapidly in character, becoming symmetric and antisymmetric
combinations of the well states.  For both A and B, the third
state has one quantum of excitation of the harmonic oscillator.
The states at $C$ have purely harmonic-oscillator character.  In
going from $A$ to $C$, the steady decrease of the well-well
separation is evident.} \label{fig3}
\end{figure}

In Fig. 3 we see that the first two eigenfunctions in the
potential of Eq. (\ref{firstjc}) change form in going from point A
to point B, from $\ket{L,R}$ to $\ket{S,A}$.  This rapid change is
actually desirable for the operation of the qubit; numerical
integrations of the time-dependent Schrodinger equation show that
we obtain high-fidelity quantum operations if $\Phi_C(t)$ is
pulsed with an overall rise time of 15 nsec, with appropriate
pulse shaping (see the discussion of visibility-limiting
mechanisms below).

Without the coupled harmonic oscillator, this $X$-rotation scheme
suffers from the problem that $\Delta$, and thus $\theta$, is
exponentially sensitive to the value of $\Phi_C$.  This
exponential sensitivity to the barrier height is evident in Fig. 2
for values of $\Phi_C$ near $B$.  However, the presence of the
oscillator degree of freedom changes several features of the
spectrum in a dramatic way.  The essentially horizontal lines in
Fig. 2 are the equally spaced oscillator energy levels, which are
insensitive to the flux applied to the qubit.  But because of the
mutual inductance $M$ in the last term of Eq. (\ref{firstjc}),
there is an avoided crossing between the oscillator and qubit
eigenlevels.  For the parameters of the device in Fig. 1, the
``vacuum Rabi splitting" $2g$ (in the notation of the
Jaynes-Cummings model, see \cite{Yale}) indicated in Fig. 2 is
about 220 MHz.  For the 15 nsec ramp-up time mentioned above, the
time-evolution remains accurately adiabatic through this
anticrossing as well. Thus, after ramping up $\Phi_C$ to point C
in Fig. 2, the system remains in a superposition of just two
states, the ground state and the first excited state which now has
almost entirely the character of a single quantum of excitation in
the harmonic oscillator. $\Delta$ in this regime is almost
completely independent of $\Phi_C$, meaning that the execution of
the $X$ rotation in this regime is almost entirely insensitive to
noise in $\Phi_C$.

While not of direct relevance to small-angle $X$ rotations that
will be desired for quantum gates, this insensitivity is very
valuable for performing the analog of a Ramsey-fringe experiment
to characterize this qubit\cite{vion}.  The protocol for this
experiment is as follows:  First, prepare the system in the state
$\ket{L}$, which in the ideal case ($\epsilon=0$) is an equal
superposition of the energy eigenstates $\ket{S}$ and $\ket{A}$.
Second, ramp up the control flux $\Phi_C$ adiabatically, then hold
it constant at a value $\Phi_{hold}$ past the anticrossing with
the harmonic oscillator state (i.e., to point $C$ in Fig. 2).
After hold time $T$, ramp down the control flux to the resting
state, then measure the qubit in the $\ket{L,R}$ basis. The
measured $L$ probability should show Larmor oscillations of the
form $\cos\omega_TT$, where
$\omega_T=1/\sqrt{C_TL_T}$\cite{foot2}.

We believe that the inaccuracy in the setting of the resting
state, i.e., the inability to set $\epsilon$ exactly equal to
zero, is likely to be the most important mechanism for reducing
visibility in our system.  If the resting state is not exactly
symmetric, then, since $\Delta$ is exponentially small at point
$A$, it is likely that $\Delta<<\epsilon$.  In this case $\ket{L}$
is not an equal superposition of energy eigenstates, but is an
eigenstate itself, as Fig. 2a shows.  Thus, a completely adiabatic
evolution will keep the system at all times in an instantaneous
eigenstate, and no Larmor oscillations will take place -- the
visibility will be zero.

The resolution to this dilemma is that, for the realistic case in
which $\epsilon\neq 0$, the time evolution in the Ramsey
experiment we have discussed is at a crucial moment actually {\em
nonadiabatic}, in a way that permits the experiment to succeed as
described.  There is a ``portal" in the $\epsilon$-$\Delta$
parameter space, which, if passed through at a finite rate,
results in a successful $X$ rotation (Ramsey-fringe visibility
near 100\%).  The idea is that since $\Delta$ rises exponentially,
$\Delta(t)=\Delta_0e^{t/\tau}$, there is a brief interval of time
during which $\Delta(t)\approx\epsilon$; this sets the location of
the portal in the parameter space.  This is the critical time
because here the form of the eigenstates of $H$ changes quite
rapidly from nearly $\ket{L,R}$ to very nearly $\ket{S,A}$, as
Fig. 3 shows.  The evolution here will be nonadiabatic if
$\epsilon\tau/\hbar$ is not too large. That is,
$-1\leq\epsilon\tau/\hbar\leq 1$ approximately defines the width
of the portal.  If the time evolution passes through the portal,
then nearly half of the state amplitude is transferred from the
ground state to the excited state, resulting in essentially 100\%
visibility.

This picture has been confirmed in detail by numerical
simulations, and also emerges from an exact solution of the
two-state problem in which $\epsilon$ is time independent and
$\Delta$ increases exponentially\cite{Req}; then the exact
solution of the time-dependent Schrodinger equation is
proportional to $c_+(t)\ket{L}+i c_-(t)\ket{R}$, where
$c_\pm(t)=\exp(t/2\tau)J_{\pm1/2+i\epsilon\tau/\hbar}[\Delta_0\tau/\hbar
\exp(t/\tau)]$, and $J_\nu(x)$ is a Bessel function.  With this
exact solution the visibility for finite $\epsilon$ can be
calculated to be ${\mbox{sech}}^2(\pi\epsilon\tau/2\hbar)$. Our
numerical simulations confirm this dependence: given the existing
precision of pulsed magnetic-flux control (a few $\mu\Phi_0$), a
nonzero $\epsilon$ in the MHz range will typically be present.
This implies that the portal will typically be located near
$\Phi_C=0.35\Phi_0$ as in Fig. 2. We numerically find a
Ramsey-fringe visibility in excess of 90\% if $\Phi_C(t)$ is
initially ramped rapidly through the portal (from $A$ to $B$ in
0.8 nsec), and then brought slowly up to its final value (from $B$
to $C$ in 15 nsec).  The necessary time dependent flux control is
demanding, but can be achieved in our lab.

We believe that our qubit and its control schemes show good
promise for scalability to larger systems.  The relatively large
size of the main loops (Fig. 1) means that establishing strong
inductive couplings between qubits is straightforward, permitting
fast two-qubit gates to be done.  As we have already demonstrated,
its large size also makes reliable, and potentially fast, readout
of the qubit state quite simple. The strong coupling to a
transmission line degree of freedom permits $X$ rotations to be
done that are very insensitive to flux noise, and, as other
workers have recently shown, the presence of a controllable
harmonic-oscillator quantum also offers new possibilities for
quantum computing architectures, including easy motion of qubits
and long-distance coupling of stationary qubits. Our control
schemes now considerably enrich this picture.

DPDV is supported in part by the NSA and ARDA through ARO contract
numbers W911NF-04-C-0098 and DAAD19-01-C-0056.  Support of DARPA
contract MDA972-01-C-0052 is also acknowledged.  We thank Fred
Brito, Guido Burkard, Dennis Newns, and the authors of \cite{Req}
for helpful discussions.


\begin{thebibliography}{100}
\bibitem{foot1} Cubic and higher order terms in $\delta$ are
necessary for an accurate quantitative description of $V_Q$, but
all the important qualitative features are captured by the simple
form Eq. (\protect\ref{firstjc}).
\end{thebibliography}
\end{document}